\documentclass[prl,aps,amssymb,showpacs,twocolumn]{revtex4}
\usepackage{amsmath}
\usepackage{amssymb}
\usepackage{amsthm}
\usepackage{amsfonts}
\usepackage{enumerate}
\usepackage{latexsym}
\usepackage{psfrag}
\usepackage{graphicx}

\newcommand{\beq}{\begin{equation}}
\newcommand{\eneq}{\end{equation}}

\input{epsf}

\begin{document}

\tolerance 10000

\newcommand{\vk}{{\bf k}}


\title{Spin Splitting and Spin Current in Strained Bulk Semiconductors}
\author { B. Andrei Bernevig and Shou-Cheng Zhang }

\affiliation{Department of Physics, Stanford University,
         Stanford, California 94305 }

\begin{abstract}
We present a theory for two recent experiments in bulk strained
semiconductors \cite{kato1, kato2} and show that a new, previously
overlooked, strain spin-orbit coupling term may play a fundamental
role. We propose simple experiments that could clarify the origin of
strain-induced spin-orbit coupling terms in inversion asymmetric
semiconductors. We predict that a uniform magnetization parallel to
the electric field will be induced in the samples studied in
\cite{kato1, kato2} for specific directions of the applied electric
field. We also propose special geometries to detect spin currents in
strained semiconductors.
\end{abstract}


\maketitle

Spin manipulation in semiconductors has seen remarkable theoretical
and experimental interest in recent years with the advent of
spin-electronics and with the realization that strong spin-orbit
coupling in certain materials can influence the transport of
carriers in so-called spintronics devices \cite{wolf}. In
particular, the issue of creating spin polarization of carriers in
nonmagnetic semiconductors with spin-orbit coupling using only
electric fields has caused a flurry of theoretical and
experimental activity \cite{murakami1, sinova, Dyakonov:1971,Hirsch:1999,Zhang:2000, Levitov:1985,Edelstein:1990,Aronov:1991,Magarill:2001,Chaplik:2002,%
Inoue:2003,Cartoixa:2001,Silsbee:2001}.
 Two kinds of theories of
spin-polarization under the action of an electric field have been
put forward. The first kind, dating back since the mid 1980's
\cite{Levitov:1985}, predicts the existence of a spatially
homogeneous net spin polarization perpendicular to the applied
electric current in two dimensional samples with spin-orbit
interaction. This effect is dissipative and has been recently
observed experimentally \cite{ganichev}. There also exist two very
recent \cite{murakami1, sinova}  theories predicting
non-dissipative, intrinsic spin currents with the spin polarization
and flow direction perpendicular to each other and to the electric
field. This effect does not create a bulk magnetization but, if
observed, can be used for spin injection, and its validity is being
experimentally tested at the present time. One of the theories
\cite{sinova} predicts a spin current polarized out of plane and
flowing perpendicular to the in-plane electric field applied on a
$2$-dimensional semiconductor sample exhibiting Rashba spin-orbit
coupling. As long as the Rashba spin splitting is large enough, the
spin conductivity is 'universal' ($e/8 \pi \hbar$) in the sense that
it does not depend on the value of the coupling. The other effect
\cite{murakami1} appears in the valence band of the bulk samples and
is proportional to the spin-orbit splitting of the valence bands (to
the difference between the Fermi momenta of the heavy and light-hole
bands).

In the first part of this letter we analyze the theory behind two
recent experiments in bulk strained semiconductors \cite{kato1,
kato2} where an electric-field-induced uniform homogeneous spin
polarization upon an applied electric field is observed. We make the
 case that the observed spin-splitting (whose origin is
puzzling) and spin polarization is due to a previously overlooked
strain-spin-splitting term, and propose easy experimental checks of
our theory.

In the second part of this letter we predict the appearance of an
intrinsic spin polarized spin current in n-doped bulk (and 2
dimensional) strained semiconductors (GaAs, GaSb, InSb, InGaAs,
AlGaAS, etc) under the influence of an electric field. The spin
conductance is 'universal', in the sense that it does not depend on
the value of strain (for large enough strain), but it is
proportional to the average Fermi momentum of the conduction band.
The effect is due to the spin-orbit splitting of the conduction band
under strain and is hence absent in strain-free semiconductors. The
very long spin relaxation time in the conduction band as well as the
relative penetration of strain engineering in semiconductor industry
applications make this effect of potential technological importance.
We propose an experimental technique using the already existing
setup in \cite{kato1, kato2} to measure the spin current and to
differentiate between the intrinsic spin current and the uniform
magnetization effects.

In \cite{kato1} nine samples of n-doped ($n=3 \times 10^{16}
cm^{-3}$) $In_xGa_{1-x} As$ ($x = 5\% - 7 \%$) of thicknesses
between $200 nm - 1500 nm$, grown in the $[001]$ direction on
undoped GaAs substrate, are used to probe the electron spin dynamics
through time and spatially resolved Faraday rotation (FR). The
length and width of the samples are roughly $300\mu m \times 80 \mu
m$. The lattice mismatch provides for diagonal strain in the $x, y,
z = [100], [010], [001]$ directions of $0.04\% - 0.46\%$
\cite{kato3} (contrary to claims in \cite{culcer}, the lattice
constants in $x$ and $y$ directions are also strained, this being a
generic feature of $[001]$ growth). Moreover, anisotropic shear
strain develops in all directions ($xy, xz, yz = [110], [101],
[011]$) due to different direction-dependent strain relaxation rates
at the growth temperature of around $500C$ \cite{kavanagh}. This
guarantees that all the components of the strain tensor
$\epsilon_{ij}, \; i,j =x,y,z$ are non-zero and of the same order of
magnitude. The magnitude of the strain components is given in
Table[1].

Pump-probe FR beams measure the total magnetization of the optically
injected electron spins in the growth direction $z$ when the samples
are placed in an electric field on the $[110]$ and $[1 \bar{1} 0]$
directions, respectively. The dynamics of the spin packet is mainly
described by a precession around a \emph{total} magnetic field
$\vec{B}_{tot} = \vec{B}_{int} + \vec{B}_{ext}$ where
$\vec{B}_{ext}$ is an externally applied magnetic field whereas
$\vec{B}_{int}$ is the momentum-dependent internal magnetic field
caused by the spin-orbit coupling. The precession around a
$\vec{B}_{int}$ is the main feature of most of the spintronics
devices, starting with the Das-Datta spin-field transistor
\cite{dasdatta}. Under an applied electric field, the average
particle momentum acquires a non-zero value, parallel to the
electric field. The internal magnetic field is caused by the spin
orbit coupling: the electric field acts on the particle momentum
which in turn couples to the spin. The signal at the probe beam can
be fitted to $\cos(g \mu_B |\vec{B}_{int} + \vec{B}_{ext}| \Delta
t/\hbar)$ where $\mu_B$ is the Bohr magneton, $g$ is the electron
$g$-factor while $\Delta t$ is the temporal delay between the pump
and probe pulses. This fit gives the direction and value of
$\vec{B}_{int}$ which turns out to be perpendicular to the applied
electric field $E$ and the $\hat{z}$ axis (for $E$ in-plane); the
value of $\vec{B}_{int}$ is used to determine the spin splitting
$\Delta_0 = g \mu_B B_{int}$ and a phenomenological relation
$\Delta_0 = \beta v_d$ is observed where $v_d$ is the spin-drift
velocity and $\beta$ is a constant of proportionality that is the
focus of the experiment \cite{kato1}. Experiments find that $v_d$ is
linearly proportional to the electric field $E$.

As a first step, let us theoretically address the question of origin
of $\beta$. By group theory, inversion symmetry breaking bulk
strained semiconductors exhibit three main types of spin splitting
\cite{pikustitkov}:
\begin{multline} \label{hamiltonians}
 H=\frac{\hbar^2}{2m} k^2 + H_1+H_2+H_3, \\
H_1 = \lambda [\sigma_x k_x (k_y^2 - k_z^2) + \sigma_y k_y
(k_z^2 - k_x^2) + \sigma_z k_z (k_x^2 - k_y^2) ] \\
H_2 = \frac{1}{2} C_3 [ \sigma_x (\epsilon_{xy} k_y - \epsilon_{xz}
k_z) + \sigma_y (\epsilon_{yz} k_z - \epsilon_{yx} k_x) + \\ +
\sigma_z(\epsilon_{zx} k_x - \epsilon_{zy} k_y) ] \\
H_3 = D[\sigma_x k_x (\epsilon_{zz} - \epsilon_{yy}) + \sigma_y k_y
(\epsilon_{xx} - \epsilon_{zz}) + \sigma_z k_z (\epsilon_{yy} -
\epsilon_{xx})]
\end{multline}
\noindent $m = 0.0665 m_0$ is the effective electron mass in the
conduction band \cite{vurgaftman}, $\lambda, C_3, D >0$ are material
constants, $\sigma_{x,y,z}$ are the 3 spin-Pauli matrices,and
$\epsilon_{ij}, \;\;  i,j =x,y,z$ are the components of the
symmetric strain tensor.

\begin{figure}
  \includegraphics[width=155mm]{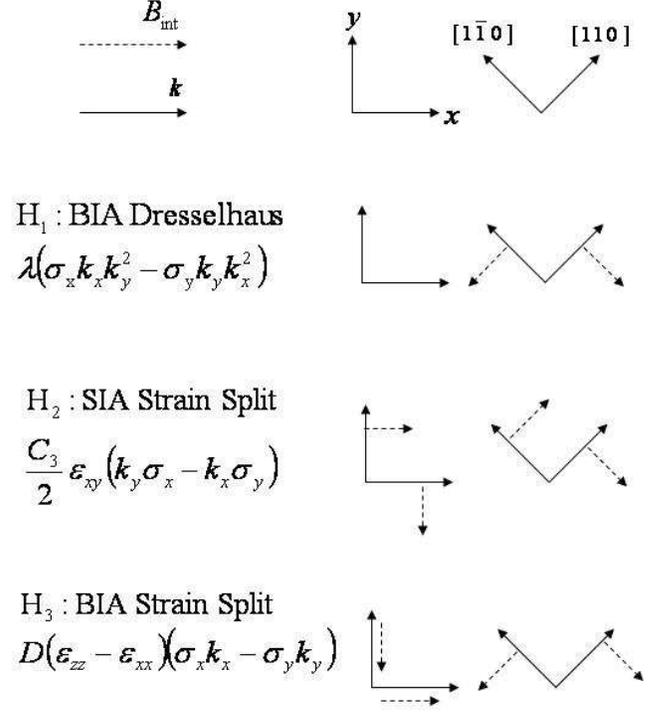}\\
  \caption{Direction of the internal magnetic field $\vec{B}_{int}$ for the three spin splitting
  Hamiltonians $H_1, H_2, H_3$, ($\lambda, C_3, D>0$) considering the electric field $E$ (and hence the average momentum) to be in
  plane. It was also assumed that $\epsilon_{xx} = \epsilon_{yy}$, as appropriate in the experiment \cite{kato1}.
  The experimental SIA data cannot be explained by the term $H_2$,
  but nevertheless, the SIA-type $B_{int}$  seen in the experiment
  has the same direction as the one plotted here. When $E|| x$ or $E||y$, the
  BIA term $H_3$ induces a $B_{int} ||E$ whereas the BIA term $H_1$ does not induce spin splitting for these directions.
  This constitutes a simple check of the experiment}\label{SIAvsBIA}
\end{figure}

All three Hamiltonians can be written as the coupling of a
fictitious $k$-dependent internal magnetic field $\vec{B}_{int}$ to
the electron spin, $\vec{B}_{int} (k) \vec{\sigma} = B_x (k)
\sigma_x +B_y (k) \sigma_y + B_z (k) \sigma_z$ (an overall factor of
$g \mu_B$ has been absorbed into the definition of $B$ to simplify
notation). The directions of $\vec{B}_{int}$ as dependent on the
directions of $\vec{k}$ are shown in Fig[\ref{SIAvsBIA}]. The
SIA-type term gives a $\vec{B}_{int}$ that keeps its orientation as
$\vec{k}$($||\vec{E}$) is rotated between the $[110]$ and
$[1\bar{1}0]$ directions, while both BIA-type $\vec{B}_{int}$ coming
from $H_1$ and $H_3$ change their sign between $[110]$ and
$[1\bar{1}0]$. The difference between $H_1$ and $H_3$ is that the
latter has a finite $\vec{B}_{int}$ when $\vec{E} || x,y$ whereas
the former has zero $B_{int}$ for the same directions.

In \cite{kato1} the values of the splitting $\beta$ are measured on
the $[110]$ and $[1\bar{1}0]$ directions and because of the
sign-changing properties of $\vec{B}_{int}$ mentioned above, the BIA
and the SIA contributions to $\beta$ can be obtained as follows:
$\beta_{BIA} = (\beta_{[110]}-\beta_{[1\bar{1}0]})/2, \; \beta_{SIA}
= (\beta_{[110]}+\beta_{[1\bar{1}0]})/2$. Surprisingly, the spin
splitting is more of a BIA-type rather than an SIA-type,
contradicting the conventional knowledge that an SIA-type term
described by $H_2$ is responsible for spin splitting in strained
semiconductors \cite{pikustitkov, hassenkam, howlett, khaetskii,
bahder, seiler}.

 Theoretically, the Dresselhaus term $H_1$ is a bulk-inversion asymmetry
term that appears even in the absence of strain. As observed in the
experiment, the fictitious internal magnetic field $\vec{B}_{int}$
is perpendicular to the momentum $\vec{k}$: $\vec{B}_{int} (k)
\vec{k} =0$, where $B_x = \lambda k_x (k_y^2 - k_z^2)$, $B_y$ and
$B_z$ being obtained by cubic permutation. For GaAs, the constant
$\lambda = 22 eV \AA^3$. However, we believe this term is not
responsible for the spin splitting observed in the experiment
\cite{kato1}. The observed splitting is linear in momentum $k$,
 inconsistent with the $H_1 \sim k^3$. Experiments performed on InSb \cite{seiler}, another material with
inversion asymmetry, support this conclusion and point strongly to
the fact that strained InSb is described by $H_2$. In \cite{seiler}
stress of up to $4$ kbar is applied mechanically on a $1\times  1
\times 10 mm^3$ sample and Shubnikov-de-Haas oscilations are used to
probe the band structure. Without applied strain, the conduction
band exhibits a spin-splitting that is small and cubic in $k$,
described by $H_1$. In \cite{seiler} the application of diagonal
does not induce any observable spin splitting whereas the
application of shear strain induces a splitting linear in $k$,
described by $H_2$. Relatively large stress-induced splitting of the
Fermi surfaces occurs in the lower concentration ($n=1.4 \times
10^{15} - 2.0 \times 10^{17} cm^{-3}$) samples \cite{seiler}. The
energy splitting dispersion switches from $k^3$ in the unstrained
case to $k$ when strain (stress) above $1kbar$ is applied, in
accordance to $H_2$ becoming dominant over $H_1$. From a theoretical
estimate, at $n=3\times 10^{16} cm^{-3}$, $H_1$ should be of the
same size as $H_3$ and roughly one order of magnitude lower than
$H_2$. The spin splitting at the Fermi wavevector $k_F= 0.96\times
10^8 m^{-1}$ due to $H_1$ is less than $ 10^{-5} eV$. By contrast
the spin splitting due to $H_2$ is $C_3 \epsilon_{xy} k_F = 5.06
\times 10^{-2} \epsilon_{xy} eV = 5\times 10^{-5} - 15 \times
10^{-5} eV$ for $\epsilon_{xy} = 0.1 \% - 0.3 \%$ as in \cite{kato1}
(see Table[1] for conversion of strain components from \cite{kato1}
to the orthogonal system $\epsilon_{xy}$). An experimental value of
$C_3 / \hbar = 8 \times 10^5 m/s$ for GaAs was used
\cite{d'yakonov}. Contrary to previous remarks \cite{culcer}, there
is hence no theoretical or experimental a-priori reason to disregard
the strain-dependent spin splitting terms in favor of the
Dresselhaus $k^3$ term for the doping values in the experiment
\cite{kato1}.

There are a number of experimental reasons in \cite{kato1} hinting
the marginal significance of the $k^3$ term. In \cite{kato1} strain
plays a critical role in generating the spin-orbit coupling
$\vec{B}_{int}$. Samples prepared from the same wafer but unstrained
show a reduction by an order of magnitude of $B_{int}$ along both
the $[110]$ and the $[1\bar{1}0]$ directions. If $H_1$ were
responsible for spin splitting, its value would remain unchanged
upon varying strain. Strain could only enter the system through the
variation of the effective electron mass in the $x,y,z$ directions,
as \cite{culcer} points out. However, these variations with strain
are of a maximum $2\% -3\%$ \cite{culcer, vurgaftman} thereby not
accounting for the order of magnitude variation of the
spin-splitting between the strained and the unstrained cases
observed in \cite{kato1}.

The term $H_2$ is a structural inversion asymmetry (SIA)-type term
that has its origin in the acoustic phonon interaction of the
valence band with the conduction band \cite{pikustitkov}. In the
framework of the Kane's $8\times 8$ matrix ($2\times2$ for the
conduction and split-off band and $4\times 4$ for the valence band)
the conduction band couples to the valence band. In systems with
inversion symmetry where the selection rules for \textbf{L} are
satisfied, it is impossible to couple spin-$0$ ($|s \rangle$) with
spin-$1$ ($|z \rangle$) through a spin-$2$ term ($\epsilon_{ij}$)
and hence $\langle s | \epsilon_{xy} | z \rangle =0$. However, when
inversion symmetry is broken, the fore-mentioned term need not be
zero as the \textbf{L} selection rule need not apply. Upon
straining, the matrix elements between the conduction and valence
band have the form $\langle s | \epsilon_{xy} | z \rangle$ (plus
cyclic permutations) where $|s \rangle$ is the $s$-orbital and $|z
\rangle$ is one of the $p$ orbitals.  Through perturbation theory,
one can compute the effect of this valence-conduction band
interaction when projected to the conduction band and obtain the
conduction band effective Hamiltonian $H_2$ \cite{pikustitkov,
howlett, khaetskii, bahder}. Taking into account that the electric
field is in-plane ($<k_z>= 0$) and that $\epsilon_{xy} \ne 0$ (see
Table[1]), in $H_2$ the components of the internal magnetic field
(which due to the rescaling by $g \mu_B$ has units of energy) are:
$B_x = \frac{1}{2} C_3 \epsilon_{xy} k_y, \; B_y  =  - \frac{1}{2}
C_3 \epsilon_{yx} k_x = - \frac{1}{2} C_3 \epsilon_{xy} k_x$.
Switching coordinates to the $[110]$ and $[1\bar{1}0]$ directions,
$B_{[110]} = \frac{1}{2} C_3 \epsilon_{xy} k_{[1\bar{1}0]} , \;
B_{[1\bar{1}0]} = -\frac{1}{2} C_3 \epsilon_{xy} k_{[110]}$ (see
Fig[\ref{SIAvsBIA}]. Since $H_2$ is an SIA term, the spin splitting
$\beta$ will be of SIA type $\beta^{th}_{SIA}$ (th stands for the
theoretical estimate). Since $<k> = \frac{1}{\hbar} m v_d$ where
$v_d$ is the drift velocity of the spin packed due to the electric
field, we get a simple formula for the
\begin{equation}
\beta^{th}_{SIA} = \frac{C_3}{\hbar} \epsilon_{xy} m
\end{equation}
\noindent By using the experimentally known value for $C_3/\hbar = 8
\times 10^{5} m/2$, the predicted values for $\beta^{th}_{SIA}$ are
given in Table \ref{straintable}. The theoretical values are larger
than the observed ones by a factor of $3-30$ and no matching trend
between the data and the SIA term $H_2$ can be found. Moreover, as
remarked in \cite{kato1} no systematic correlation between the
experimentally observed SIA contribution and the strain is observed.
We hence come to the conclusion that the SIA spin splitting observed
in \cite{kato1} is not induced by the uniform shear strain (which
would give the values $\beta_{SIA}^{th}$ and which have been
confirmed in mechanical experiments) but borrows substantially from
the dislocations and strain gradient inherent in growing such a
thick sample through MBE techniques. This is not to say that the SIA
term is negligible: as seen in Table \ref{straintable} the SIA term
is substantial and comparable in magnitude with the BIA term.
However, the SIA term does not correlate with strain ad cannot be
described by $H_2$.

\begin{table}
  \centering
\begin{tabular}{|c|c|c|c|c|c|c|c|c|}
  \hline
   Sample &  $ \epsilon_{zz}$  &  $\epsilon_{xx}=\epsilon_{yy}$  &  $ \epsilon_{xy}$  & $\beta^{exp}_{SIA}$ &  $\beta^{th}_{SIA}$ & $\beta^{exp}_{BIA}$ & $\frac{\beta^{exp}_{BIA}}{\epsilon_{zz} - \epsilon_{xx}}$ & $D/\hbar$ \\
  \hline
   A  & 0.46 & -0.16 & 0.2 & -24 & 604 & 75 & 121 & 1.59\\
  \hline
   B & 0.14 & -0.2 & 0.08 &  -26 & 241 & 13 & 38 & 0.5 \\
  \hline
   E & 0.13 & -0.42 & 0.18 &  69 & 543 & 43 & 78 & 1.03\\
  \hline
   F & 0.07 & -0.32 & 0.12 & 54 & 362 &31 & 79  & 1.04\\
  \hline
   G & 0.04 & -0.32 & -0.04 &  44 & -121 & 31 & 86 & 1.13\\
  \hline
   H & 0.13 & -0.42 & 0.26 &   -2 & 785 & 24 & 43 & 0.56\\
  \hline
   I & 0.04 & -0.16 & 0.2 &  65 & 604 & 23& 115 & 1.51\\
  \hline
\end{tabular}
 \caption{Strain components (in $\%$) in \cite{kato1} converted to
  cartesian coordinates (C and D are compositionally graded films \cite{kato1}).
  The conversion equations, obtained after a simple coordinate transformation, are:
  $\epsilon_{zz} =\epsilon_{[001]}, \epsilon_{xx} =\epsilon_{yy} = \frac{1}{2} (\epsilon_{[110]} + \epsilon_{[1\bar{1}0]}),
  \epsilon_{xy} = \frac{1}{2}(\epsilon_{[110]} -\epsilon_{[1\bar{1}0]}) $. The measured spin splitting slope
  values $\beta^{exp}_{SIA}$ and $\beta^{exp}_{BIA}$ as well as the
  theoretical value $\beta^{th}_{SIA}$
  obtained using the Hamiltonians $H_2$ for the SIA - type splitting are given in $neV \; nS \; \mu
  m^{-1}$, the units of  \cite{kato1}. The value of the deformation
  potential $D$ in the term $H_3$, determined as explained in
  Eq(\ref{valueofD}) is given for each sample. The discrepancies between the values of
  $\beta^{exp}_{SIA}$ and $\beta^{th}_{SIA}$ make the case that
  $H_2$ is not the full mechanism for strain SIA splitting. For five (A, E, F, G, I) out of the seven
   samples the values of the
deformation potential $D$
  deduced from the $H_3$ model of the strain BIA splitting are
  within $30\%$ of each other. Combined with the simplicity of the
  model, this match gives weight to the suggestion that BIA
  splitting is produced by a $BIA$ term. $\beta^{exp}_{BIA}/(\epsilon_{zz}
  -\epsilon_{xx})$ is measured in $10^2 neV \; nS \; \mu
  m^{-1}$, while $D$ is measured in $10^4 m/s$
    }\label{straintable}
\end{table}

The remaining spin splitting term is $H_3$. Although this term is
allowed by group theory, it only shows up at higher order in
perturbation theory than $H_2$ in the $k \cdot p$ method. We claim
that in the experiment \cite{kato1} this term is responsible for the
spin-splitting observed, and determine the value of the constant
$D$. We note, however, that this does not settle the theoretical
puzzle of why the $H_3$ would be more significant than $H_2$ in this
case, which might have to do with the conditions of the experiment
such as low temperature, the appearance of dislocations and strain
gradient. $H_3$ is a Hamiltonian of BIA-type and vanishes with
vanishing strain, thereby satisfying two of the experimental
observations in \cite{kato1}. From Eq.[\ref{hamiltonians}], the
internal magnetic field $\vec{B}_{int}$ reads: $B_x =
Dk_x(\epsilon_{zz} - \epsilon_{yy}),\;\; B_y = Dk_y(\epsilon_{xx} -
\epsilon_{xx}),\;\; B_z = Dk_z(\epsilon_{yy} - \epsilon_{xx})$, and
since the in-plane electric field influences only the in-plane
momentum, $<k_z> =0$ and only the two in-plane components of the
internal magnetic field remain. In accordance with the experiment,
we place $\epsilon_{xx} =\epsilon_{yy}$, and $\epsilon_{zz} -
\epsilon_{xx}>0$ (Table[\ref{straintable}]). We hence have  $B_x =
k_x D (\epsilon_{zz} - \epsilon_{xx} ),\;\; B_y = k_y
D(\epsilon_{xx} - \epsilon_{zz})$, $D (\epsilon_{zz} - \epsilon_{xx}
)>0$. Since $\vec{B}_{int}$ is not perpendicular to $\vec{k} ||
\vec{E}$: $B_x k_x + B_y k_y = ( k_x^2 - k_y^2)D (\epsilon_{zz} -
\epsilon_{xx} )$, one may think this term is incompatible with the
observed $\vec{B}_{int} \perp \vec{E} || \vec{k}$ in \cite{kato1}.
This, however, would be hasty: the experiment is performed in only
two directions, with $\vec{E} ||[110]$ and $\vec{E} ||[1\bar{1}0]$,
for which $k_x = \pm k_y$. For these two directions only, the
$\vec{B}_{int}$ in $H_3$ is perpendicular to the momentum and the
electric field, hence satisfying a major constraint the experimental
data poses on the theory. Since the value of the constant $D$ is
unknown from previous experimental studies (although it was
suggested that they can be sometimes sizable \cite{rocca}) there is
no way of theoretically predicting the values of the spin-splitting
from our model. However, we can check if the model is consistent
with the experimental data and we can also obtain a value of the
constant $D$ which, being a material constant, should be similar on
all the samples cited here. Since $<k> = \frac{m}{\hbar} v_d$ where
$v_d$ is the spin drift velocity along the spin packet we find:
\begin{equation}
\beta_{[BIA]}^{th} = 2 \frac{D}{\hbar} (\epsilon_{zz} -
\epsilon_{xx} ) m
\end{equation}
\noindent We can determine the value of $D$ from the experimental
data for $\beta$ and strain $\epsilon$:
\begin{equation} \label{valueofD}
\frac{D}{\hbar} = \frac{1}{2m}
\frac{\beta_{[BIA]}^{exp}}{\epsilon_{zz} - \epsilon_{xx} }
\end{equation}
\noindent As a consistency check, since $D$ is a material constant,
${\beta_{[BIA]}^{exp}}/({\epsilon_{zz} - \epsilon_{xx} })$ should be
quasi-constant between the samples quoted in the experiment. In 5
out of the 7 samples studied in \cite{kato1}, the values of
${\beta_{[BIA]}^{exp}}/({\epsilon_{zz} - \epsilon_{xx} })$ are close
together to within $30\%$, lumped in two groups (samples $A,I$ are
very close to each other, and within $30\%$ of the value for $E,F,G$
which are again very close between themselves). The samples $E,F,G$
were grown in the same day. The deviant samples $B,H$ were also
grown in the same day, and hence the variation of the constant
coefficient $D$ within a sample set that was grown on the same day
is less than $15\%$ \cite{kato5}. Different growth conditions are
most likely responsible for the (still small) variations between
samples grown in different days. The consistency check is further
proof that $H_3$ is the term responsible to the spin-splitting in
\cite{kato1}. The values obtained for D are given in
Table[\ref{straintable}].

We showed that $H_3$ is a BIA-type Hamiltonian vanishing with
vanishing strain, with an internal magnetic field that is
perpendicular to the applied electric field for the two experimental
directions $[110]$ and $[1\bar{1}0]$ and which is consistent with
the reported data for the spin splitting. On the other hand, $H_1$
and $H_2$, the previously known spin splitting terms, fail to
reproduce the data on more than several counts.It is easy to
experimentally prove, using the setup in \cite{kato1}, that $H_3$ is
responsible for the spin splitting is easy: one would measure the
internal magnetic field due to BIA on the x or y direction. In this
case, an $H_3$ term would give an internal magnetic field parallel
to $\vec{E}$ (of course, there will also be an internal $\vec{B}$
from an SIA term that is still perpendicular to $\vec{E}$, but a
component of $\vec{B}_{int}$ parrallel to the electric field should
be easily detectable).

In another beautiful experiment, Kato \emph{et al.} measure through
Farraday Rotation (FR) a nonzero uniform magnetization $\rho_{el}$
induced by driving an electric current (electric field) through the
sample E of their previous experiment \cite{kato1}. It has been long
predicted \cite{Levitov:1985, Edelstein:1990} that semiconductors
with spin-orbit coupling will exhibit a uniform magnetization when
placed in an electric field generating a charge current. This can be
trivially understood by a simple argument: writing the spin-orbit
Hamiltonian as a $\vec{k}$ dependent magnetic field Zeeman coupled
to spin, $\vec{B}_{int} (k) \vec{\sigma}(k)$, the application of an
electric field $\vec{E}$ will make the average value of the momentum
be non-zero $<\vec{k}> = \frac{e}{m} \vec{E} \tau$ where $\tau$ is
the momentum relaxation time. This creates a non-zero average
$<\vec{B}_{int}> = \vec{B}(<\vec{k}>)$ which orients the spins along
its direction through the Zeeman-like coupling of the spin-orbit
term.

 We now try to numerically estimate the value of the
 uniform magnetization $\rho_{el}$ using the BIA-type $H_3$.
  From \cite{kato2}, the BIA contribution to the uniform magnetization
can be obtained as $\rho_{el}^{BIA} =\frac{1}{2} (\rho_{el}^{\vec{E}
|| [1\bar{1}0]} - \rho_{el}^{\vec{E} || [110]})$ and is around $3
\times 10^{18} m^{-3}$ for $E =10^4 V/m$. We will now try to
estimate this from first principles using $H_3$ as the main BIA term
and using the value of $D$ for sample $E$ deduced in
Table[\ref{straintable}]. A simple linear response calculation of
the magnetization $\sigma_i$ to the electric current $J_j$ (due to
the applied electric field $E_j$) gives:
\begin{multline}
\\
\rho_{el}^i = \frac{2 \pi e \tau}{\hbar} Q_{ij}  E_j  \\
Q_{ij} = \langle T \sigma_i J_j \rangle = \int \frac{d^3 k}{(2
\pi)^3} \frac{n_{E_-} - n_{E_+}}{B^2} \left( B_i \frac{\partial
B}{\partial k_j} - B \frac{\partial B_i}{\partial k_j} \right)
\end{multline}
\noindent where $i,j =x,y,z$ and $B_i(k)$ are the components of the
internal magnetic field for $H_3$, $B =\sqrt{ \sum_{i=x,y,z} B_i
B_i}$, $n_{E_\pm}$ is the Fermi function of the spin-split energies
$E_{\pm} = \frac{\hbar^2}{2m} k^2 \pm B$ for $H_3$. For the
Hamiltonian $H_3$, considering $\epsilon_{xx} =  \epsilon_{yy}$ we
obtain for:
\begin{equation}
 \rho_{el}^{BIA} = \frac{e\tau m k_F D(\epsilon_{zz} -
\epsilon_{yy})}{\pi \hbar^3} E
\end{equation}
\noindent  As previously pointed out the magnetization is parallel
to the electric field for $\vec{E} ||\hat{x}$ or $\vec{E} ||
\hat{y}$. This provides an important and easy check of the above
assumption that the observed strain spin splitting comes from $H_3$.
The only two directions where $\vec{\rho}_{el}$ is perpendicular to
the electric field are $[110]$ and $[1\bar{1}0]$, the directions on
which the experiment is performed. Considering a sample of mobility
$\mu =0.6 m^2/Vs$ \cite{kato3} we obtain an estimate for $\rho_{el}
= 3.45 \times 10^{18} m^{-3}$ for a field $E=10^{4} V/m$, compared
to an experimental value of $3\times 10^{18} m^{-3}$ for the same
value of the electric field. The theoretical value obtained is
within the experiment's error margins.

\begin{figure}
  \includegraphics[width=90mm]{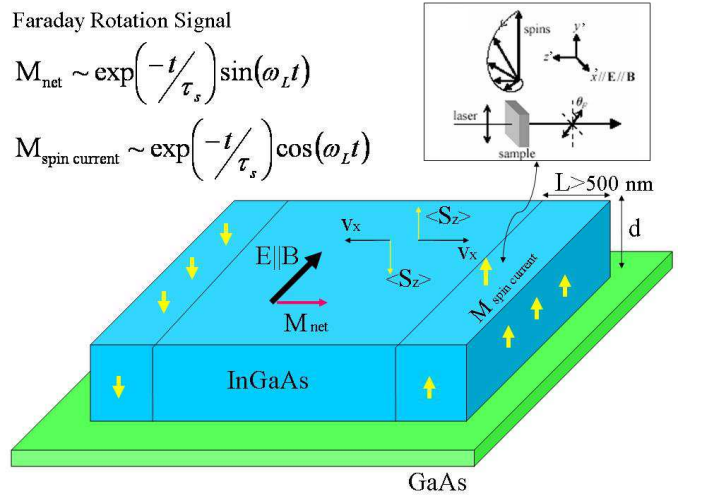}\\
  \caption{An electric field will cause both an observed net uniform bulk magnetization and a proposed spin current.
  The spin current will accumulate at the edges over a spin diffusion length of more than $500nm$ thus making its detection
  practical with a beam slightly more focused than in \cite{kato2}. The uniform magnetization and the spin current spin accumulatio
  are $\pi/2$ out of phase. }\label{spincurr}
\end{figure}

Finally, using the current setup in \cite{kato1, kato2} we propose
an experiment to test the prediction of dissipationless spin
current. For spin 1/2 two-dimensional systems, the initial
prediction \cite{sinova} is subject to some sort of controversy,
\cite{Inoue:2003, halperin, nomura} as the introduction of
impurities apparently makes the spin current vanish. We here adopt
the alternative view and propose a clear-cut experiment which can
see the spin accumulation due to the spin current. Similar to the 2D
case, in the present case, the application of an electric field
$E_j$ to a semiconductor with spin orbit coupling will create a spin
current $S_l^i$ flowing perpendicular to the electric field and
polarized perpendicular to both the field and the direction of flow.
Using linear response, the expression for the spin conductance is:
\begin{multline}
\\ J_i^l = \sigma_{ij}^l E_j \\
\sigma_{ij}^l = \frac{\hbar^2 }{2m}\int \frac{d^3 k}{(2 \pi)^3}
\frac{n_{E_-} - n_{E_+}}{B^3} k_i \epsilon_{lnm} B_n \frac{\partial
B_m}{\partial k_j}
\end{multline}
\noindent where $i,j,l,m,n =x,y,z$, $\epsilon_{lnm}$ is the totally
antisymmetric tensor in 3 dimensions and $B_i(k)$ are the components
of the internal magnetic field. For $H_3$ and for $\epsilon_{xx}
=\epsilon_{yy}$ the only non-zero components of the spin conductance
are:
\begin{multline}
\sigma^3_{21} = - \sigma^3_{12}= \frac{e}{\hbar} \frac{1}{D
(\epsilon_{zz} - \epsilon_{yy})}\frac{\hbar^2}{m} \int
\frac{d^3k}{(2 \pi)^3}
\frac{n_{E_-} - n_{E_+}}{(k_x^2 + k_y^2)^{3/2}} k^2_y = \\
=\frac{e}{\hbar} \frac{1}{D (\epsilon_{zz} - \epsilon_{yy})
}\frac{\hbar^2}{2m} \frac{1}{ (2 \pi)^3} \int_{0}^{2\pi} d\phi
\int_0^\pi d\theta \sin^2{\phi} (k_-^2 - k_+^2)
\end{multline}
\noindent where $\phi, \theta$ are the polar angles of
$\overrightarrow{k}$ and where $k_-, k_+$ are the fermi momenta of
the two bands. When both bands are occupied (positive Fermi energy),
we find $k_- - k_+ = \frac{2 m}{\hbar^2} \frac{\lambda(k)}{k}, \;
\lambda(k) = D (\epsilon_{zz} - \epsilon_{xx}) k \sin \theta$.
Usually the spin splitting is much smaller than the Fermi energy,
and we can define an average Fermi momentum $k_F = \frac{1}{2} (k_-
+ k_+) \approx (3 \pi^2 n)^{1/3}$, $n$ being the dopant density.
With this, we find that the spin-conductivity will be independent of
the value of the strain:
\begin{equation}
 \sigma^3_{21} = -
\sigma^3_{12} = \frac{e}{\hbar} \frac{k_f}{4 \pi^2}
\end{equation}
\noindent
 The
result for the spin conductance is intermediate between the 2D spin
1/2 spin current and the 3D spin 3/2 spin current. Similar to
\cite{murakami1} but unlike  \cite{sinova} the spin conductance
depends on the fermi momentum, a characteristic of the 3D. Unlike
\cite{murakami1}, but similar to \cite{sinova}, the spin conductance
does not depend on the strength of spin-orbit coupling. Even though
the spin conductance does not depend on the value of strain, it is
essential that spin-orbit splitting (due to strain in this case) be
present. Upon the application of an electric field on the $x$ axis,
a spin current will flow on the $y$ axis spin polarized in the $z$
direction. For $n=10^{16} cm^{-3}$, and a field $E =10^4 V/m$ we
estimate a spin current $j_{spin} = \frac{e}{\hbar} \frac{k_f}{4
\pi^2} E = 2 \times 10^{21} (\mu_B/ cm^2 s)$ where $\mu_B$ is a Bohr
magneton. Since spin conductivity varies as $n^{\frac{1}{3}}$ and
charge conductivity varies as $n$, for low values of $n$ the spin
conductance will overtake the charge conductance and the spin
current will be larger than the charge current caused by the
electric field. The density at which this happens is $n^{2/3} <
\frac{2 e}{\hbar \mu} \frac{(6 \pi^2)^{1/3}}{8 \pi^2}$, where $\mu$
is the mobility in the sample, or $n< 2 \times 10^{16} cm^{-3}$ for
a sample of mobility $\mu=0.6 m^2/Vs$.

 The
flow will result in accumulation on the opposite $zx$ faces of the
crystal (see Fig[\ref{spincurr}]). For the present experiment, we
estimate this spin accumulation of the order of $J_{spin} \tau_S =
10^{13} \mu_B /cm^2$. Due to the extremely spin life time of above
$1 ns$, the distance from the edge of the sample, the spin diffusion
length is very large, of the order $L=500 nm -1 \mu m$. The FR beam
used in \cite{kato2} has a resolution of $4.7 \mu m / 9.7 \mu_m$ on
the x and y axis respectively, but focusing the beam within $1\mu m$
is possible \cite{kato4, stephens}. Then, if the spin current
prediction is right, applying the FR beam on the edge of the sample
should give a clear signal (larger than the uniform magnetization in
the bulk). Since the uniform magnetization and the spin accumulation
due to spin hall current are perpendicular to each other, in
time-resolved FR experiments, the spin current spin accumulation and
the uniform magnetization are out of phase by $\pi/2$ (see
Fig[\ref{spincurr})

In conclusion, we have analyzed two very recent experiments
\cite{kato1, kato2} and proved that conventional spin splitting
terms and strain spin splitting terms do not explain the data. We
have introduced a previously largely unknown term and made the case
as to why it explains the observed features in \cite{kato1, kato2}.
We have proposed further simple experiments to verify our
assertions. If true, our proposal gives rise to the clear
possibility of obtaining a uniform magnetization parallel to the
applied electric field, as opposed to the one perpendicular to it
that has been observed so far. Along with predicting a 3D spin
current, we have also proposed a way to test the spin currents in
spin 1/2 systems.

BAB wishes to primarily thank Y. Kato for many essential discussions
and explanations regarding the experiments \cite{kato1, kato2}. The
authors also wish to thank H. Manoharan for many stimulating
discussions on strain-related issues and E. Mukamel for critical
discussions relating to the experiments \cite{kato1, kato2}. Many
thanks also go to G. Zeltzer, L. Souza De Mattos for discussions on
strain growth and H.D. Chen for useful technical support. B.A.B.
acknowledges support from the Stanford Graduate Fellowship Program.
This work is supported by the NSF under grant numbers DMR-0342832
and the US Department of Energy, Office of Basic Energy Sciences
under contract DE-AC03-76SF00515.

\end{document}